\documentclass[referee]{raa}            

\newcommand{\orcid}[1]{%
    \raisebox{0.7ex}{\scalebox{1}{
        \href{https://orcid.org/#1}{\includegraphics[height=1.5ex]{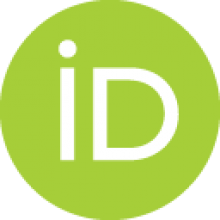}}%
    }}%
}
\usepackage{graphicx,times}             
\usepackage{natbib}
\usepackage{amssymb,amsmath}
\usepackage[utf8]{inputenc}
\usepackage{textgreek}
\bibpunct{(}{)}{;}{a}{}{,}
\usepackage[pagebackref=true]{hyperref}

\begin{document}

  \title{SVOM/VT: Real-Time Onboard Data Processing
}

   \volnopage{Vol.0 (202x) No.0, 000--000}      
   \setcounter{page}{1}          

\author{Hong-Bo Cai
      \inst{1,*}\footnotetext{$*$Corresponding Authors, these authors contributed equally to this work.}
    \and Yu-Lei Qiu\orcid{0009-0007-7207-4884}
    \inst{1,*}
    \and Li-Ping Xin\orcid{0000-0002-9422-3437}
    \inst{1}
    \and Zheng-Yang Bian
    \inst{2}
    \and Rui-Feng Su
    \inst{2}
     \and Qing-Yun Mao
    \inst{2}
    \and Bin-Ping Su
    \inst{2}
    \and Jun-Wang He
    \inst{2}
    \and Wei Gao
    \inst{3}
    \and Jian Zhang
    \inst{3}     
    \and Li-Jun Dan
    \inst{3}
    \and Kun Chen
    \inst{2}
    \and Dong Li
    \inst{2}
    \and Chao Wu\orcid{0009-0001-7024-3863}
    \inst{1}
    \and Hua-Li Li
    \inst{1}
    \and Jin-Song Deng\orcid{000-0001-5646-8583}
    \inst{1,5}
    \and Yong-He Zhang
    \inst{2}
    \and Jian-Yan Wei
    \inst{1}  
   \and Bertrand Cordier
      \inst{4} 
   }

   \institute{National Astronomical Observatories, Chinese Academy of Sciences,
             Beijing 100101, China; {\it chb@nao.cas.cn, qiuyl@nao.cas.cn}\\             
        \and
                Innovation Academy for Microsatellites, Chinese Academy of Science, Shanghai 201203, China;\\
        \and
                         Xi’an Institute of Optics and Precision Mechanics, Chinese Academy of Sciences, Xi’an 710119, China; \\    
        \and
             CEA/Paris-Saclay, Irfu/Département d'Astrophysique, 91191 Gif-sur-Yvette, France;\\
        \and  School of Astronomy and Space Science, University of Chinese Academy of Sciences, Beijing 100101, China\\ 
 \vs\no
   {\small Received 202x month day; accepted 202x month day}}

\abstract{ 
The \textit{SVOM} Visible Telescope (VT) is critical for the rapid identification of gamma-ray burst (GRB) optical counterparts, particularly for high-redshift candidates that require immediate infrared spectroscopic follow-up. To address the stringent bandwidth constraints of the VHF downlink while ensuring real-time data availability, we developed the VT Onboard Data Processing Pipeline (VOPP). This paper details the software architecture, algorithms, and hardware implementation of VOPP using an FPGA and a CPU. The pipeline performs essential real-time tasks, including image quality assessment, dark and flat-field correction, and optimized image stacking to mitigate cosmic ray contamination and variable background noise. Furthermore, it generates compact source catalogs and highly compressed 1-bit images to facilitate rapid downlink. In-flight performance analysis confirms the pipeline’s robustness, demonstrating the availability of VT VHF data for 78\% of promptly slewed \textit{SVOM} GRBs, with 56\% leading to the identification of optical counterparts, typically within 18 minutes post-trigger.
}

   \authorrunning{H. Cai, Y. Qiu, L. Xin, et al.}            
   \titlerunning{Real-Time Onboard Data Processing}  

   \maketitle

%
%
\section{Introduction}           
\label{sect:intro}


The Space-based Variable Objects Monitor (SVOM) satellite \citep{2015arXiv151203323C,2016arXiv161006892W}, a joint China-France mission, is dedicated to detecting, localizing, and studying gamma-ray bursts (GRBs)—the most luminous transient phenomena in the universe—along with other high-energy transients through multi-wavelength observations. 
Gamma-ray bursts (GRBs) are widely believed to originate from two primary mechanisms: the gravitational collapse of massive stars \citep{2006ARA&A..44..507W}, which generates long GRBs (LGRBs) capable of probing the early Universe up to redshifts of ( z $\approx$ 20 ) \citep{2000ApJ...536....1L}, and the merger of compact objects (e.g., neutron star binaries or neutron star–black hole systems), whose relativistic jets power short GRBs (SGRBs) \citep{2006RPPh...69.2259M,2015PhR...561....1K}. The landmark association of GRB 170817A with the gravitational wave event GW170817 \citep{2017PhRvL.119p1101A} not only confirmed neutron star mergers as progenitors of SGRBs but also inaugurated the era of multi-messenger astronomy, unifying electromagnetic, gravitational-wave, and neutrino observations.

The VT is one of the four scientific payloads onboard the \textit{SVOM} satellite \citep{Cordier+etal+2026}, alongside ECLAIRs (a wide-field gamma-ray imager), GRM (non-imaging wide-field gamma-ray monitors), and MXT (a narrow-field X-ray telescope). The VT's primary objective is to rapidly respond to GRB triggers, originating either from \textit{SVOM} or other missions such as \textit{Swift} \citep{2004ApJ...611.1005G} or the \textit{Einstein Probe} \citep{2025SCPMA..6839501Y}, to detect their optical counterparts \citep{vt_overview_qiu+etal+2026}. The telescope features a 44 cm aperture Ritchey-Chrétien system operating at f/8 and utilizes a dichroic beamsplitter for simultaneous imaging in two bands (400–650 nm and 650–1000 nm). In comparison to the \textit{Swift} UVOT payload (170–650 nm) \citep{2005SSRv..120...95R}, the VT's sensitivity extends up to 1000 nm, a range optimized for the preselection of high-redshift GRB candidates. 

The VT is designed to downlink preliminary quick-look data via the VHF network \citep{Cordier++etal+vhf+2026}, a system enabling rapid dissemination of \textit{SVOM} GRB information. This capability facilitates the rapid identification of optical counterparts, enhancing GRB localization accuracy, and guiding large ground-based telescopes to measure redshifts before the optical afterglow becomes too faint. Even a non-detection (upper limit) by the VT is highly valuable, as it may suggest a potential high-redshift GRB candidate, thereby mobilizing large telescopes with near-infrared capabilities for follow-up observations. This strategy’s efficacy is exemplified by the discovery of the high-redshift GRB 250314A ($z=7.3$) \citep{2025arXiv250718783C}.

To enable rapid data delivery, we developed the VT Onboard Data Processing Pipeline (VOPP), which generates compact quick-look products for downlink. This pipeline performs critical tasks such as flat-field correction, source extraction, and catalog generation, transmitting compact data products instead of full frames. Because single-exposure images are susceptible to cosmic-ray contamination, onboard stacking is employed to enhance sensitivity and mitigate artifacts. Furthermore, earthshine-induced stray light complicates observations, necessitating real-time image quality checks to reject heavily contaminated frames. Although the bandwidth and speed of the VHF system are lower than those of the TDRSS(Tracking and Data Relay Satellite System) used by \textit{Swift}, and the GRB localization error box produced by MXT is larger than that of \textit{Swift}'s X-ray telescope, these measures ensure reliable VT data delivery within \textit{SVOM}’s telemetry limits without compromising scientific performance.

In this paper, we present the software design of the VOPP, detailing its main workflow (Section \ref{sect:strategy}), onboard image quality evaluation (Section \ref{sect:quality}), calibration methods (Section \ref{sect:calibration}), the image processing pipeline (Section \ref{sect:processing}), and the performance test of VOPP (Section \ref{sect:performance}). We conclude by summarizing the system’s capabilities and outlining potential future enhancements (Section \ref{sect:summary}).


\section{workflow of VOPP}

\label{sect:strategy}

\begin{figure}
        \centering
        \includegraphics[width=0.9\linewidth]{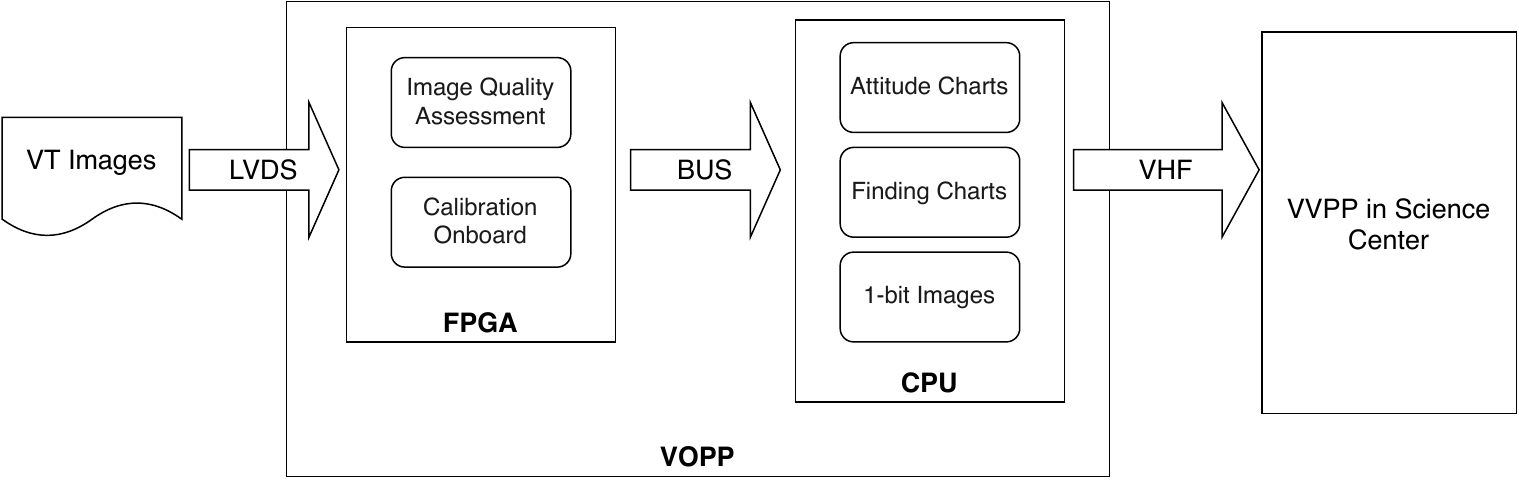}
        \caption{Schematic flowchart of the VT Onboard Data Processing Pipeline.}
    \label{fig:VT_PDPU}
\end{figure}

Figure \ref{fig:VT_PDPU} illustrates the schematic flowchart of the VOPP. The pipeline is executed by the Payload Data Processing Unit (PDPU), which is primarily composed of a million-gate Field-Programmable Gate Array (FPGA) and an Atmel 697F CPU (60 MIPS). In addition to VT data processing, the PDPU manages other critical functions, including the implementation of pointing strategies and VHF control. Given these competing demands, the computational resources available for VT data processing are limited, necessitating a carefully designed and efficient onboard processing strategy.

\begin{figure*}
    \centering
     \includegraphics[width=0.9\linewidth]{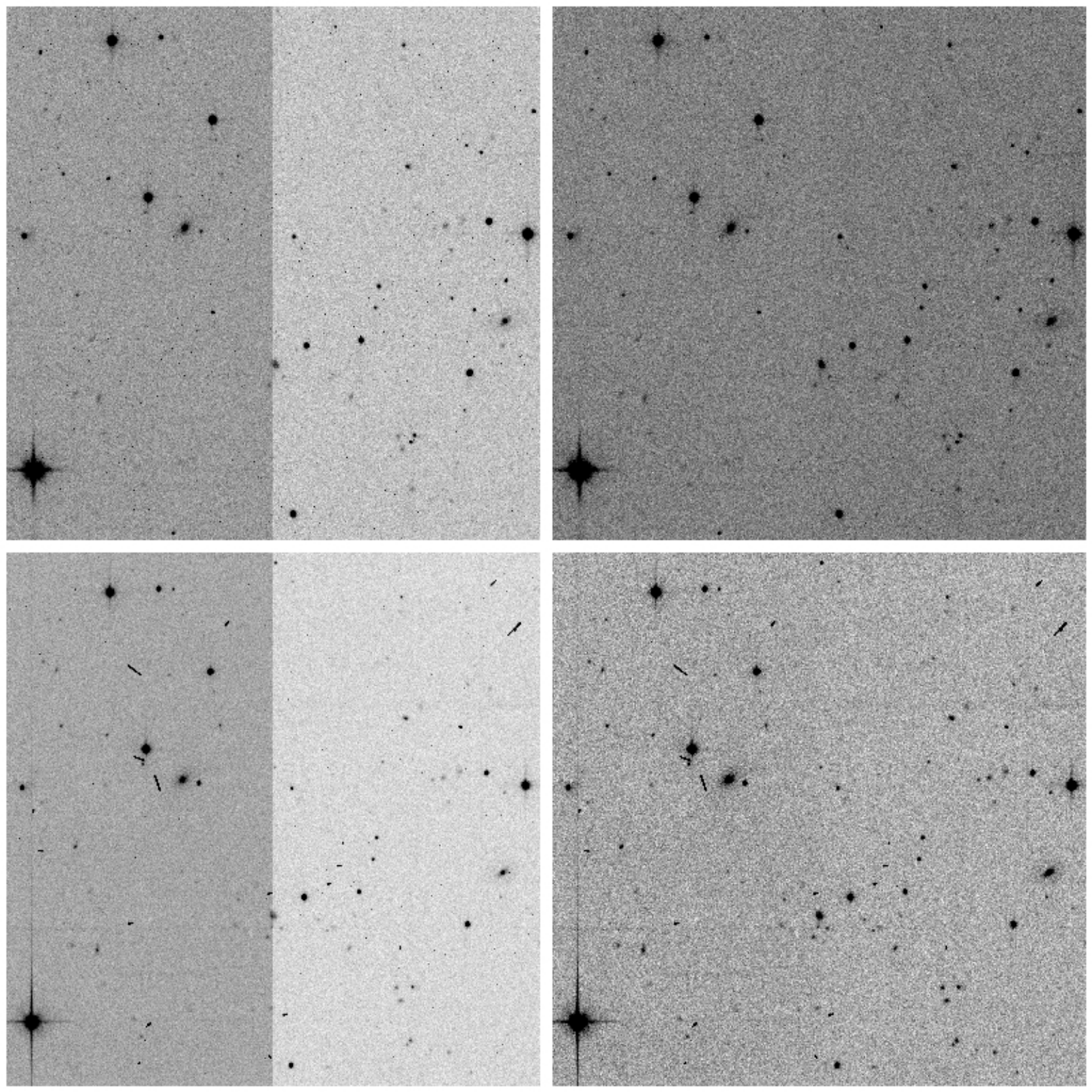}
    \caption{Comparison of original and calibrated 400 × 400-pixel image sections from the GRB 250314A field. Top: Blue-band raw (left) and calibrated (right) images. Bottom: Red-band raw (left) and calibrated (right) images.}
    \label{fig:orig_cali}
\end{figure*}


\textbf{FPGA-Based  Calibration and Quality Assessment:} FPGAs excel at high-speed serial data processing, making them ideal for pixel-by-pixel operations in imaging applications, such as dark-frame and flat-field corrections. Their parallel architecture enables efficient real-time processing, which is critical for handling large datasets with minimal latency.

To ensure high-quality onboard data processing, it is essential to pre-select optimal images, specifically those characterized by low stray light (minimal background noise) and high platform stability (yielding optimal PSF performance). The FPGA executes these assessments in real time using computationally efficient algorithms, as detailed in subsequent sections.


\textbf{Image Processing with CPU:} The CPU performs the onboard image processing, which entails sophisticated algorithms such as the processing of Attitude Charts (Section \ref{sect:ATC}), Finding Charts (Section \ref{sect:FDC}), and 1-bit Images (Section \ref{sect:1bit}). These tasks are inherently suited for general-purpose computing. Moreover, implementing these operations on a CPU simplifies software updates relative to FPGA-based solutions.

The results, including Attitude Charts, Finding Charts, and 1-bit Images, are downlinked to the ground science center via VHF, where final GRB extraction and localization are completed by the VVPP (VT-VHF data Processing Pipeline).

\section{Image quality assessment onboard}
\label{sect:quality}

The VT image quality is determined by two primary factors: platform stability and background noise levels.

\subsection {Assessment of Platform Stability}

We use the platform’s stability during the exposure as a proxy for VT’s point spread function (PSF), which is a key factor in evaluating image quality. Directly measuring the PSF on the FPGA is complex and resource‑intensive, while the PSF in long‑exposure images is strongly influenced by platform stability.

The Attitude and Orbit Control System (AOCS) \citep{lidong+etal+2026} calculates stability performance using prior data and transmits it to the VT every second. The VT imaging software logs a binary stability indicator (1 if stability meets the specified threshold, 0 otherwise) each second and records the cumulative count in the image header. The FPGA subsequently checks this count: if it falls below a predefined threshold, the image is discarded; otherwise, the image advances to background assessment.

\subsection{Assessment of Image Background}
\label{sect:assessment_background}

Accurate assessment of the VT background is essential for rejecting images affected by high background levels caused by Earthshine stray light and ensuring the selection of high-quality data. While VT operates optimally within Earth’s shadow, stray light levels rise sharply upon shadow exit. This sudden increase elevates the image background and degrades detection sensitivity due to excessive noise.

To mitigate these effects, the background is estimated using five predefined windows, deriving statistics from selected pixels while excluding bright stars to prevent contamination. Although iterative clipping is typically employed to remove stellar contributions, FPGA complexity constraints necessitates a simpler strategy: applying a fixed flux threshold and including only pixels below this limit in the statistical sample. This method demonstrates high efficacy, particularly in low star-density regions (such as high Galactic latitudes) — typical for most \textit{SVOM} GRBs — where the estimated background converges closely to the true value.

In addition to stray light, the dominant background component in VT images is zodiacal light, which peaks near the Sun and along the ecliptic plane. While \textit{SVOM}’s anti-solar pointing strategy minimizes the overall background level, the random sky distribution of GRBs introduces unavoidable fluctuations, increasing the complexity of image selection.

To account for these variations, the background selection criteria for images employ a conservative threshold, accounting for margins in \textit{SVOM} pointing uncertainties and instrumental effects (e.g., dark current). Stray light is negligible within Earth’s shadow but increases sharply upon exit, causing significant background variations between adjacent images during the transition. This sharp contrast allows for the easy exclusion of images strongly affected by stray light. However, some faint stray light may remain in the selected images due to the conservative threshold.

Consequently, most sequences (see Sec. \ref{sect:seq}) comprise 1–6 high-quality images taken entirely within Earth’s shadow. Only a small subset—typically those obtained near the shadow boundary—exhibit residual stray light. This contamination can compromise cosmic-ray rejection in image stacking. Further discussion of the issue and mitigation approaches is provided in Section \ref{sect:combination}.

\section{VT onboard calibration}
\label{sect:calibration}

The onboard calibration process involves generating master bias, dark, and flat-field images by combining individual calibration frames, maintaining the same $2048 \times 2048$ pixel format as science observations to correct instrumental effects.  Given the limited bandwidth of the S-band link and the need for frequent calibrations to track detector evolution, uplinking large processed calibration images from the ground is impractical. Consequently, both the acquisition of raw calibration data and the creation of master calibration images must be performed entirely onboard.

\subsection{ Calibration Image Generation Algorithm}

Due to CPU memory constraints, we generate only two categories of calibration images: dark frames and flat-fields. Dark frames match the exposure time of observational images, accounting for both dark current and bias. Flat-field images are acquired using LED lamps at 527 nm and 747 nm for the blue and red bands, respectively. Calibrations are performed monthly or weekly via ground telecommands.

\subsubsection{Dark images}

With the shutter closed, a set of six dark frames is acquired at long exposures (the exposure time matching that of VT observations, e.g., typical exposures of 50 s or 100 s). To generate the master dark image, we employ a median combination algorithm. This approach effectively balances cosmic ray contamination removal with signal-to-noise ratio optimization. The final master dark is saved as a 16-bit file in the PDPU memory.
  
\subsubsection{Flat-field images}

The process begins by generating a master dark image from six dark frames (typically 5 s each). Subsequently, a master flat-field image is derived from six LED flat-field frames, each with a 5 s integration time. During these exposures, the LEDs are pulsed for 2–5 ms to account for response time characteristics. Each individual flat frame is dark-subtracted using the master dark and then combined via median stacking to create a master flat-field for each VT band.

Since the image is read out through two ports, it is split into two equal parts with different gains. The processing pipeline handles this in three steps: First, the right part is scaled by a gain ratio (derived onboard or via telecommand) to match the left part. Second, the image is normalized. Third, the scaling is inverted by dividing the right part by the same ratio. This re-introduction of the gain difference is essential for the proper correction of raw observational data. The final normalized image is then scaled by 32,768 and stored as a 16-bit integer to balance precision with storage efficiency.

\subsection{Real-time Image Calibration}

During GRB observations, VT images are calibrated in real time using the FPGA. As the data stream is received, each pixel is processed using pre-loaded master dark and flat-field images. The calibration pipeline involves two stages: subtraction of the corresponding dark-frame pixel value followed by division by the normalized flat-field value. To accommodate the line-interlaced data structure of the blue and red bands, the FPGA identifies the band for each line and applies the appropriate calibration coefficients. Subsequently, the calibrated data are de-interlaced into individual single-band frames. Prior to SDRAM storage, a constant offset of 512 is added to eliminate negative pixel values, and the output is converted to integer format to optimize memory usage.
\begin{figure}
    \centering
    \includegraphics[width=0.9\linewidth]{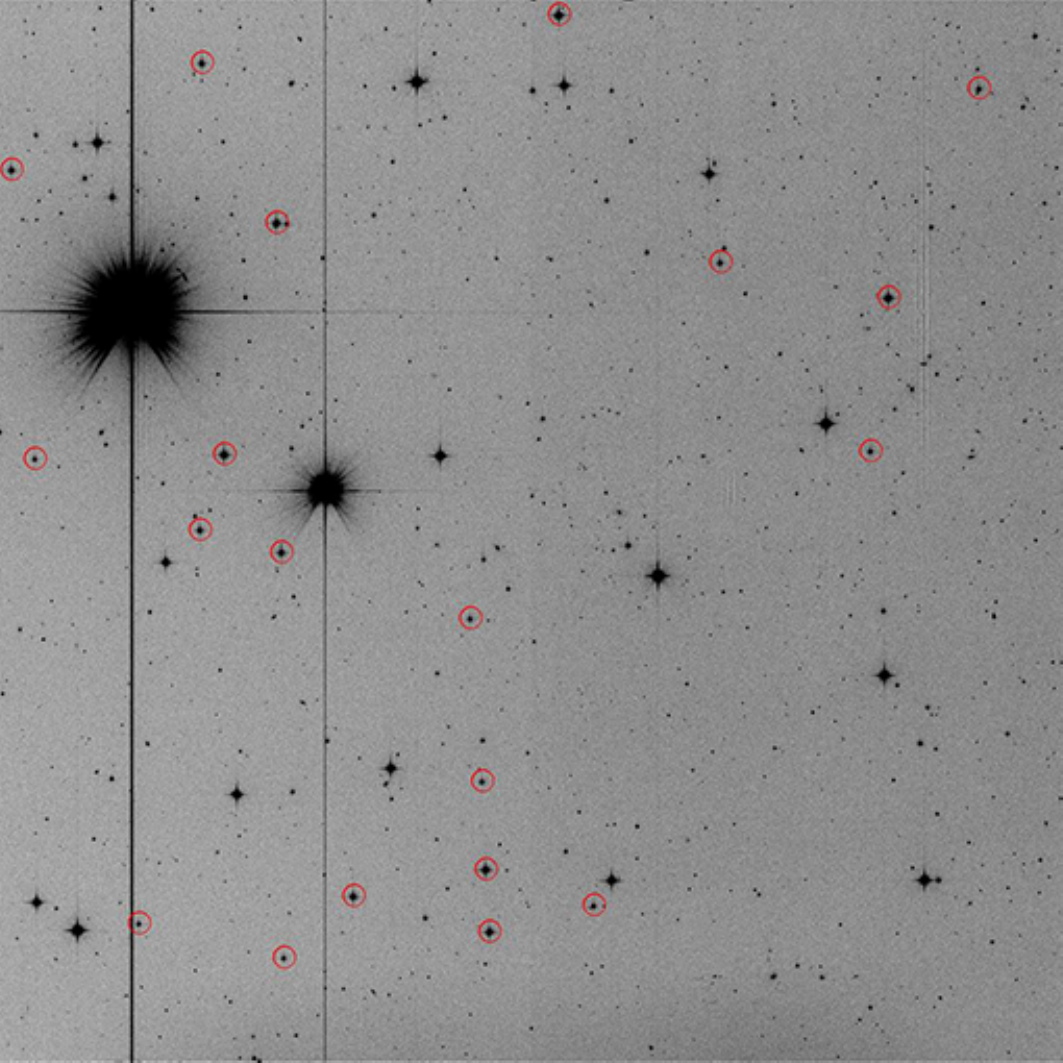}
    \caption{400 × 400-pixel sections comparing the original and calibrated images. The upper panels show the blue-band images: raw (left) and calibrated (right). The lower panels display the red-band field images of GRB 250314A.}
    \label{fig:atc}
\end{figure}
Figures \ref{fig:orig_cali}  presents 400 × 400-pixel sections of the raw and calibrated blue- and red-band field images of GRB 250314A. The calibrated images demonstrate successful suppression of most hot pixels while retaining persistent cosmic-ray events. Notably, cosmic rays in the blue band appear identical to point sources, whereas in the red band they display distinctive elongated ("worm-like") morphologies. This morphological difference stems from charge diffusion effects, which vary with substrate thickness: the blue-band CCD employs a 10 μm substrate, while the red-band CCD uses a significantly thicker 40 μm deep-depletion design. These results verify the FPGA's capability to perform robust real-time corrections while maintaining astrophysical signal integrity.

\section{image processing}
\label{sect:processing}

After onboard image calibration, the data processing sequences are executed. For each sequence, the PDPU generates three types of data products: attitude charts, finding charts, and 1-bit images. Below, we first introduce the sequences, followed by a detailed explanation of the data products.
\subsection {Sequences}
\label{sect:seq}
To optimize limited onboard resources and identify optical counterparts through their fast variability, onboard data processing is applied only to selected images. This process involves four sequences of VT continuous observations, each comprising 1 to 6 consecutive images, covering first two post-trigger orbits.

The first sequence begins once the platform stabilizes after the slew and background conditions are met. The second sequence follows immediately, capturing consecutive images. The third and fourth sequences are acquired at the start and end of the second orbit, respectively.

The number of images processed per sequence (ranging from 1 to 6) depends on the availability of high-quality images before Earth occultation in the current orbit, as onboard systems evaluate image quality based on background levels and platform stability, selecting only those meeting the criteria. Observations may also be interrupted by the South Atlantic Anomaly (SAA), potentially terminating image collection prematurely. If a sequence contains zero images—indicating no valid acquisitions—the system skips that sequence and proceeds to the next one.
\subsection{Attitude charts}
\label{sect:ATC}
Each Attitude Chart (ATC) is a data table containing the pixel coordinates of 21 bright, unsaturated stars extracted from observed images using the fast algorithm described in \citet{2010SCPMA..53S..51W}. The ATCs are downlinked to the ground via a VHF datalink channel for immediate processing. Figure \ref{fig:atc} shows an example of the 21 selected stars marked on a full-frame VT image.

On the ground, the pixel coordinates of these stars are matched to their J2000 reference positions using an astrometric catalog (e.g., Gaia). The spacecraft's attitude (pointing direction in J2000) is then determined through further processing \citep{wu+etal+2026}.

In addition to performing astrometry for sources in the VHF data, ATCs also improve the localization of MXT sources—similar to how \textit{Swift}/XRT achieves enhanced localization using \textit{Swift}/UVOT astrometry \citep{2008AIPC.1000..539E}.

\begin{figure*}
    \centering
    \includegraphics[width=0.5\linewidth]{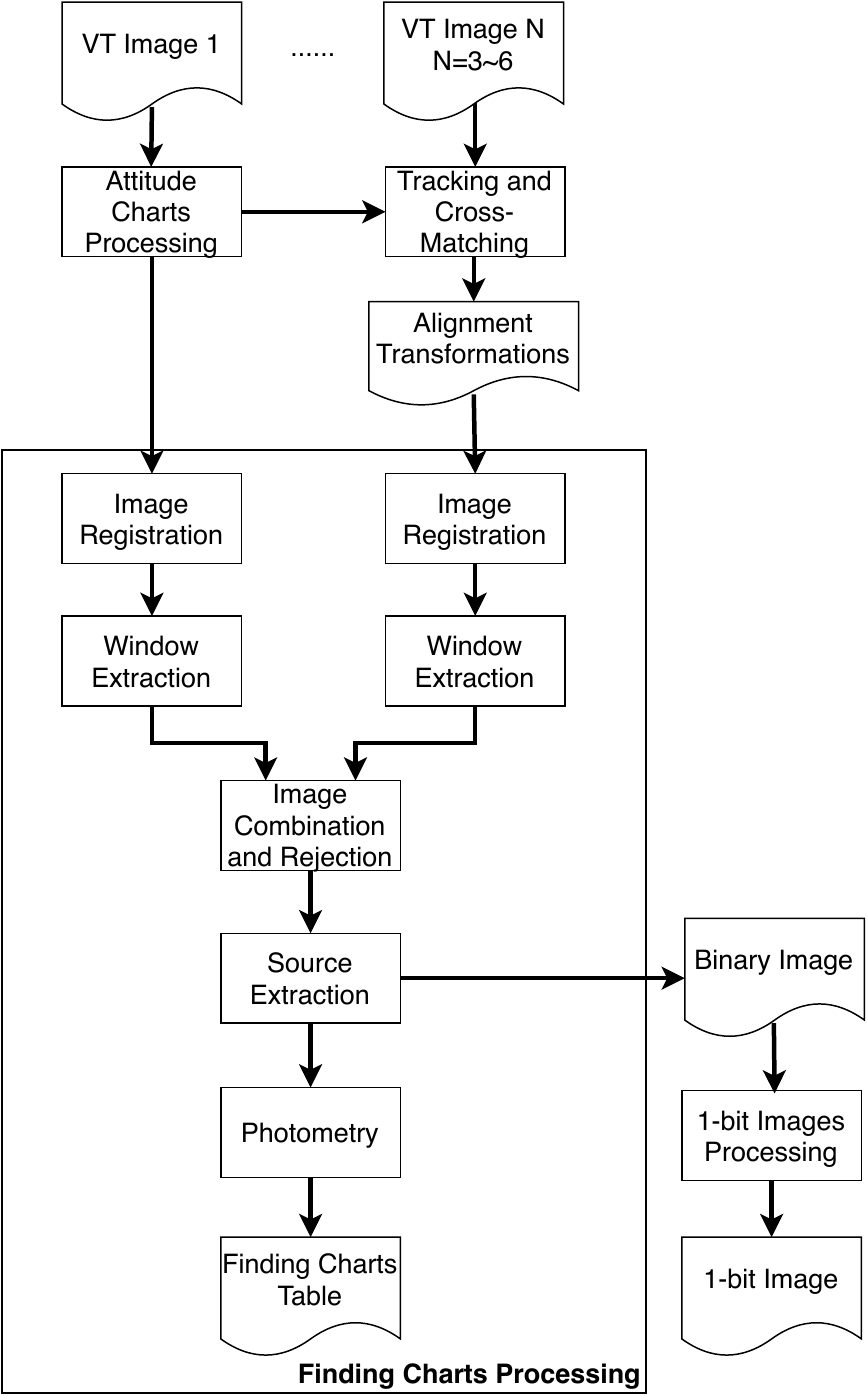}
    \caption{Imaging processing flowchart for the finding chart.}
    \label{fig:flowchart}
\end{figure*}


\begin{figure*}
    \centering
    \includegraphics[width=0.9\linewidth]{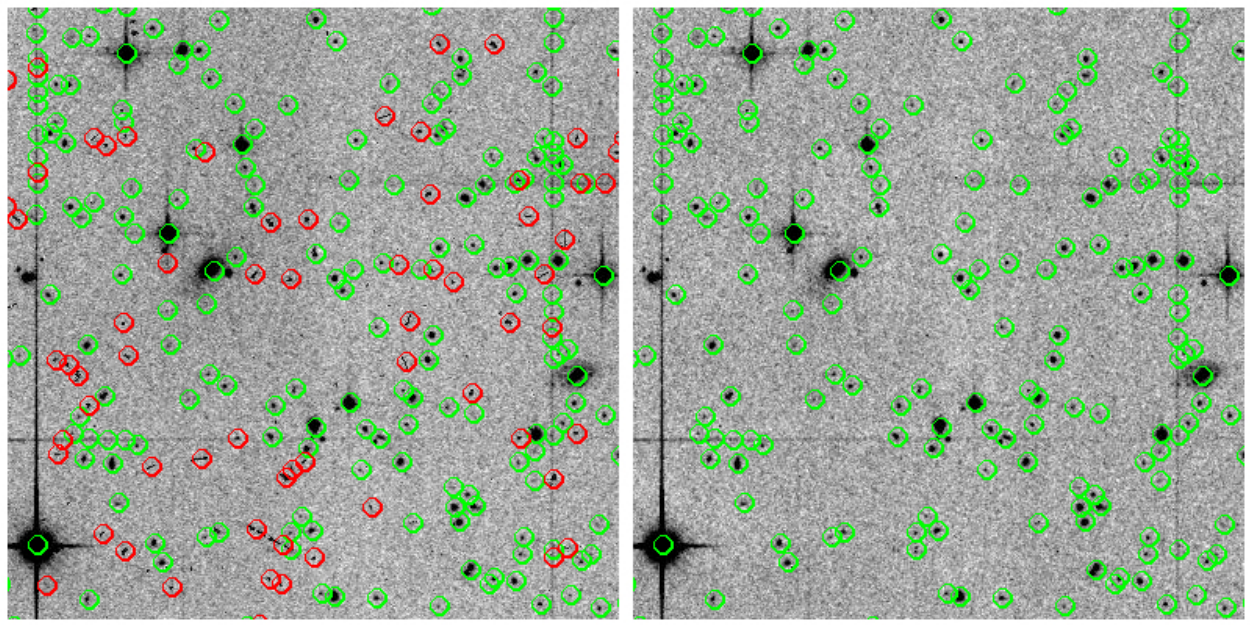}
    \caption{Comparison between the standard (left) and optimized (right) algorithms for a 400 × 400-pixel section of the GRB 250314A field image, with identified sources marked by green circles and cosmic ray events by red circles. }
    \label{fig:algorithm_combination}
\end{figure*}

\subsection{Finding charts}
\label{sect:FDC}
Finding charts (FDCs), the primary products of the VOPP, provide source lists extracted from stacked sub-images centered on MXT localizations. The sub-image size is determined to fully cover the MXT localization error region. These lists contain sources detected above a predefined signal-to-noise (S/N) threshold (currently set to 5), determined by balancing the VHF data budget and detection depth. By stacking 3–6 frames, the FDCs mitigate cosmic ray contamination and enhance the S/N of the individual sources. For each detected source in the FDCs, three aperture photometry measurements are generated (see Sec. \ref{sect:photometry}), along with ellipticity values describing the source's morphology. These parameters enable the on-ground pipelines \citep{wu+etal+2026} to identify and flag blended sources, thereby minimizing false candidates.

Below, we outline the FDC generation flowchart, followed by detailed descriptions of the image stacking and photometry algorithms.

\subsubsection {Imaging processing flowchart}
\label{sect:1bit}
In the GRB observation mode, VT performs Finding Charts and 1-bit images processing for each sequence (defined in Sec. \ref{sect:seq}). As shown in Figure \ref{fig:flowchart}, the pipeline begins with image registration. Although the platform exhibits high stability under Fine Guidance System (FGS) control (as demonstrated  performance tests in \cite{vt_overview_qiu+etal+2026} ), we apply registration universally to ensure robustness. For each sequence, only the ADC of the first image are processed. The unsaturated bright stars detected in the ADC data serve as reference points for image registration. We track and cross-match their centroids across subsequent frames to derive the alignment transformations. These transformations are then applied to co-register all subsequent images to the reference frame (the first image in the sequence).

After registration, a window is extracted from each frame, with the window center at the GRB alert position provided by the MXT at the time of the first frame’s acquisition, and the window size determined by the GRB alert localization error.

Following window extraction, the image combinations are performed. To ensure compatibility across diverse backgrounds, we developed an optimized algorithm, as detailed in Section \ref{sect:combination}.

After image combination, the workflow proceeds to source extraction. We employ an algorithm similar to SExtractor \citep{1996A&AS..117..393B} to model the background and detect sources above a predefined threshold. The binary images of detected sources serve as input for 1-bit image processing (see Sec. \ref{sec:1-bit_imges} ). 

Finally, aperture photometry is performed on the extracted sources. Further details on this process are provided in Section \ref{sect:photometry}.

 \subsubsection{Image stacking algorithm optimized for cosmic ray rejection}
\label{sect:combination}

The primary objective of image combination is to enhance the signal-to-noise ratio while mitigating cosmic ray contamination. These transient artifacts present a significant challenge in processing space telescope images due to their higher frequency compared to ground-based observations. Traditional methods typically involve stacking multiple images and rejecting outlier pixels based on a sigma threshold. However, VT images often retain residual stray light even after initial quality screening (see Sec. \ref{sect:assessment_background}). In the presence of such background gradients, direct combination proves ineffective for removing weak cosmic ray events, as their flux levels become indistinguishable from background fluctuations.

To overcome this limitation, we refine the stacking process by first subtracting a low-spatial-frequency background model from each exposure before combination. This step removes large-scale gradients—such as scattered light or uneven illumination—while preserving the intrinsic signal-to-noise ratio of astronomical sources. The background-subtracted images are then aligned and co-added into a master image. Crucially, this approach suppresses cosmic-ray artifacts without degrading image quality, as the stable background ensures cosmic rays remain clearly distinguishable. As shown in Figure \ref{fig:algorithm_combination}, the method produces significantly cleaner results compared to direct stacking. By optimizing the combination process, we achieve robust cosmic-ray rejection, reduce spurious detections, and improve source identification efficiency.

\subsubsection{Photometry algorithm}
\label{sect:photometry}

The final processing stage involves photometry of extracted sources from stacked (master) images. While most sources are point-like, the data also contain extended sources, blended stars, and residual artifacts (e.g., diffraction spikes, hot pixels, or imperfectly removed cosmic rays). These artifacts often exhibit sharp flux gradients, making morphological characterization critical for distinguishing real point sources from contaminants and reducing false transient detections.

To address this, we employ a multi-aperture photometry strategy:
(1) a small aperture (1–2 pixels in diameter) sensitive to compact cores; (2) an intermediate aperture enclosing 70\%–80\% of the total energy, optimized to maximize the signal-to-noise ratio  for point-like sources; and (3) a large aperture (or isophotal photometry), which facilitates the identification of extended sources when compared with measurements from smaller apertures.

For each master image, aperture corrections are computed and recorded in the VHF packets of the FDCs. These corrections account for the magnitude difference between the optimal (medium) aperture (2–3 pixel radius) and the full aperture (typically 5 pixel radius), calibrated using bright, unsaturated stars. This ensures that all detected sources—including GRB optical counterparts—are properly corrected to their full-aperture instrumental magnitudes.

The on-ground VT-VHF data processing pipeline (VVPP) \citep{wu+etal+2026} performs two primary tasks:
\begin{enumerate}

\item \textbf{Photometric calibration}: Converts instrumental magnitudes to the AB magnitude system through magnitude zero-point corrections determined from (1) pre-launch ground-based throughput calibration measurements and (2) periodic observations of spectrophotometric standard stars during operations (see \citet{Yao+etal+2026} for details).

\item  \textbf{Morphological classification}: Determines source morphology by comparing small$/$large and medium$/$large aperture magnitude differences against calibrated thresholds from bright, isolated stars. Objects are flagged as point-like, extended, or suspected artifacts (e.g., hot pixels or cosmic-ray residuals).

\end{enumerate}
\subsection{1-bit images}
\label{sec:1-bit_imges}
\begin{figure}
    \centering
    \includegraphics[width=0.8\linewidth]{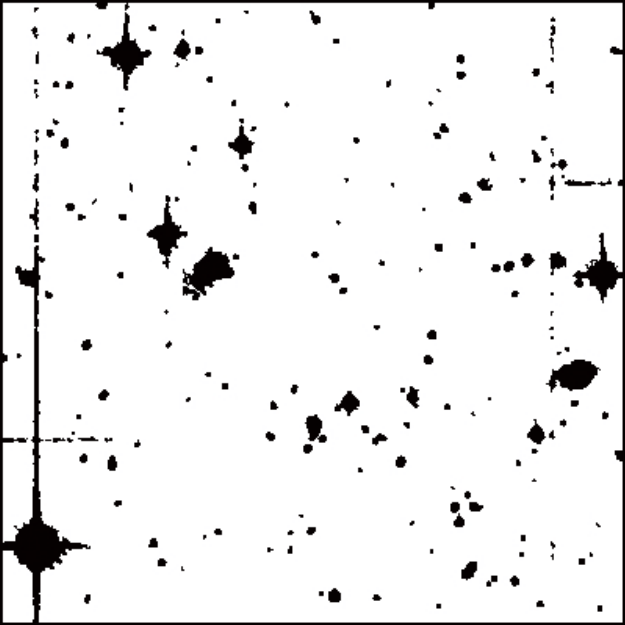}
    \caption{A 400×400-pixel 1-bit image of the GRB 250314A field.}
    \label{fig:1bitimage}
\end{figure}

Before the complete raw images are transmitted to the ground via X-band antennas, visual inspection is necessary to distinguish genuine bright optical counterparts from artifacts. While some morphological data are recorded through multi-aperture photometry, direct visual checks remain indispensable. However, downlinking full-resolution images via the VHF data channel is impractical, making it essential to simplify them to fit within the VHF bandwidth constraints.

During image processing for FDCs, a valuable by-product is the generation of 1-bit binary images. Here, each pixel is assigned a value of 0 or 1 based on whether its flux exceeds a detection threshold. This approach efficiently distinguishes true astronomical sources from artifacts while preserving key morphological features—such as point-like or extended sources and bright star refractions—despite the drastic reduction in data volume compared to the original 16-bit (or 32-bit) images.

To further reduce transmission data volume, we apply Run-Length Encoding (RLE), a lossless compression method optimized for 1-bit images. RLE encodes consecutive sequences of identical bits (0s or 1s) into compact 8-bit or 16-bit segments. Each segment consists of the bit value (encoded in the first bit) followed by its repetition count (stored in the remaining 7 or 15 bits). 

This digitization and compression strategy reduces the image size by nearly two orders of magnitude compared to the original 16-bit data, enabling efficient transmission of low-flux-resolution images while retaining essential morphological information.

To demonstrate the utility of a 1-bit image, we present Figure \ref{fig:1bitimage}, which was derived from the master image (right panel of Fig. \ref{fig:algorithm_combination}). This simplified 1-bit image preserves the morphological structure of the sources, allowing for clearer visual discrimination between genuine objects and spurious detections caused by saturation from very bright stars, bright source refraction, or blending of nearby sources. The 1-bit image significantly reduces false candidates and, when combined with the source lists (FDC), improves the efficiency of optical counterpart identification.

\section{VOPP Performance Test}
\label{sect:performance}
\begin{table*}[htbp]
    \centering
    \caption{Pre-launch test of Source consumption}
    \label{tab:prelaunch_test}
    \begin{tabular}{|l|l|r|}
        \hline
        \textbf{Processing Module} & \textbf{Sub-module} & \textbf{Average Processing Time} \\ 
        \hline
        FPGA-Based Calibration and Quality Assessment & --- & synchronized with readout \\ 
        \hline
        Attitude Charts & --- & 40 s \\ 
        \hline
        {Finding Charts} 
            & Image registration & 32 s \\ 
            & Image combination  & 24 s \\ 
            & Window extraction  & 1 s \\ 
            & Source extraction  & 33 s \\ 
            & Photometry         & 10 s \\ 
        \hline
    \end{tabular}
\end{table*}

Performance assessments were conducted both pre-launch and in-orbit. Pre-launch testing included detailed measurements of FPGA and CPU resource utilization, whereas in-orbit operations recorded only total processing time. As processing time scales linearly with the number of pixels in windowed images, all results were normalized to a standard 6 arcmin × 6 arcmin field of view. 

\subsection{Pre-lauch Performance}

The VT onboard data processing system is primarily implemented using a 3-million-gate FPGA (Xilinx XC2V300) and an Atmel AT697F CPU. The resource utilization and runtime for each product are summarized as follows:
\subsubsection{Hardware consumption}

\textbf {FPGA Utilization}: The design occupies 34\% of the available LUTs (Look-Up Tables, with a total of 3840 units) and 78\% of the available FFs (Flip-Flops, with a total of 3840 units), leaving sufficient remaining resources for potential functional upgrades and margin.

\textbf{CPU Load}:
The CPU maintains an average load of ~32\%, peaking at ~71\% during onboard data processing—well within safe operational limits.

\textbf{SDRAM Memory Footprint}:
The system operates with a fixed 360 MB pre-allocated memory footprint, including EDAC(Error Detection And Correction), eliminating dynamic allocation during runtime. Consequently, both average and peak memory usage remain stable and predictable.

\subsubsection{Runtime for Data Processing}

We focus here on the data processing pipelines for ATC and FDC, noting that 1-bit subimages are generated as by-products of FDC.

Table \ref{tab:prelaunch_test} summarizes the processing times for key production steps. The total runtime is ~40 s for ATC and ~173 s for FDC. Notably, source extraction and photometry—the only steps sensitive to stellar density—account for just 25\% of the total runtime. Since most GRBs occur at high Galactic latitudes (where stellar density is low), field-to-field variations in processing time are negligible. Thus, these results are representative of the vast majority of GRBs detected by \textit{SVOM}.

\subsection{Software Performance In-orbit }

We assessed the in-orbit performance of VOPP using data from GRB250806A. The data processing times were ~30 s for ATC and ~170 s for FDC, consistent with pre-launch test averages of 40 s (ATC) and 173 s (FDC).

\subsection{System Performance In-orbit}

The first VT VHF dataset reaches the Science Center within 18 minutes of an ECLAIRS trigger, comprising 5 min for platform slew/stabilization, 5 min exposure, ~3 min (170 s) onboard processing, and 5 min VHF transmission; subsequent sequences (typically delivered within 40 minutes) maintain similar processing efficiency despite variable VHF downlink durations due to priority scheduling. Compared to X-band’s 2–6+ hour latency, VOPP’s sub-hour response enables critical early identification of bright GRBs during peak optical luminosity—providing a $<$ 1-hour window for high-value spectroscopic follow-up with large ground-based telescopes.

As of 3 September 2025, \textit{SVOM}/ECLAIRS had detected 32 GRBs with prompt slew observations, with 25 (78\%) successfully processed on board \citep{vt_overview_qiu+etal+2026, wu+etal+2026}. Optical counterparts were identified for 14 bursts (56\% detection rate), though performance is expected to improve as both the platform and VT transition from commissioning to full operational capability.

The short latency and high success rate of VT’s VOPP solidify its critical role in early identification for \textit{SVOM} GRBs. If no optical counterpart is detected within the X-ray afterglow’s error circle, the event likely represents either a high-redshift transient or a heavily obscured burst. Early follow-up with large telescopes equipped with near-infrared facilities is essential for uncovering z \textgreater 6 GRBs, as demonstrated by the discovery of GRB 250314A at z = 7.3 \citep{2025arXiv250718783C,vt_overview_qiu+etal+2026}—the third-highest spectroscopically confirmed GRB to date. Crucially, the VOPP system’s non-detection in VHF channels guided ground-based telescopes to prioritize deep NIR observations, proving instrumental in this high-redshift breakthrough.


\section{Summary}
\label{sect:summary}

Early detection of GRB optical counterparts is critical for enabling timely spectroscopic follow-up while the afterglow remains bright, particularly for high-redshift GRBs where early optical constraints can guide subsequent near-infrared observations. The VT’s rapid response, deep-detection capabilities—with sensitivity extending to 1 μm and an optical counterpart detection rate of 70\%–80\% \citep{vt_overview_qiu+etal+2026,Li+etal+2026}—make it uniquely suited for this task. Its Onboard Data Processing Pipeline further enhances efficiency by generating quick-look results and downlinking them via a rapid VHF connection, enabling early identification of bright GRBs and optimized follow-up during the critical initial phases. The pipeline's effectiveness is underscored by its critical role in discovering the highest-redshift GRB observed in 12 years (z = 7.3) \citep{2025arXiv250718783C}.

This paper presents the pipeline's hardware architecture, functional components, and data processing flow, including specialized algorithms for VHF product generation. Analysis of VHF data from VOPP by VT ground pipelines (VVPP and VTAC, \citealt{wu+etal+2026}) shows that 78.2\% of promptly slewed GRBs have VT VHF data, with 56\% of them yielding optical counterparts—typically identified within 18 minutes post-trigger. These results confirm VOPP’s operational efficacy and highlight its potential to uncover additional high-redshift GRBs, significantly advancing GRB research.

\begin{acknowledgements}
The  \textit{SVOM} is a joint Chinese-French mission led by the Chinese National Space Administration (CNSA), the French Space Agency (CNES), and the Chinese Academy of Sciences (CAS). We gratefully acknowledge the unwavering support of NSSC, IAMCAS, XIOPM, NAOC, IHEP, CNES, CEA, and CNRS. This work is supported by the Strategic Priority Research Program of the Chinese Academy of Sciences (Grant No.XDB0550401), and by the National Natural Science Foundation of China (grant Nos. 12494571 and 12494570, 12494573, 12133003). The authors are thankful for support from the National Key R\&D Program of China (grant Nos. 2024YFA161170* and 2024YFA1611700). 
\end{acknowledgements}

\bibliography{ms2026_0023}{}

\label{lastpage}

\end{document}